\documentclass{ws-ijmpe}
\usepackage[T1]{fontenc}

\def\nuc#1#2{$^{#1}$#2}

\begin{document}

\markboth{L. Pr\'ochniak}{%
Collective states of even-even Mo isotopes}

\catchline{}{}{}{}{}

\title{\MakeUppercase{%
Microscopic study of collective states of even-even Molybdenum isotopes}
}

\author{\footnotesize \MakeUppercase{L. Pr\'ochniak}}

\address{Institute of Physics, Maria Curie-Sk{\l}odowska University,
 pl. M.Curie-Sk{\l}odowskiej~1,\\ 20--031 Lublin, Poland}

\maketitle

\begin{history}
\received{(received date)}
\revised{(revised date)}
\end{history}

\begin{abstract}
Low energy quadrupole excitations of the \nuc{84-110}{Mo} nuclei are studied
in the frame of the general Bohr collective model. The potential energy and
inertial functions are calculated using the ATDHFB method starting from the
Skyrme effective interaction. Obtained energies of the $2_1$, $4_1$, $2_2$ and $0_2$ levels and
B(E2) values for the $2_1\rightarrow 0_{\rm g.s}$ transitions are in good
agreement with experimental data. 
\end{abstract}

\section{Introduction}

The aim of this work is to describe low energy quadrupole collective
excitations of the even-even \nuc{84-110}{Mo} isotopes within the frame of
a microscopic theory based on the Skyrme effective interaction.  
We present results of calculations for $2^+_1$, $4^+_1$, $0^+_2$ and $2^+_2$
excited levels and B(E2) probability for the $2^+_1\rightarrow 0_{\rm g.s}$
transition in the whole chain of isotopes. Moreover, we include more
detailed data
on levels and E2 transitions for the \nuc{100}{Mo} nucleus.
To study the
collective phenomena the standard mean field theory must be extended with
the Adiabatic Time Dependent HFB method, which is used in this paper, or the
Generator Coordinate Method.  The important feature of the approach
presented below is that the collective space contains all quadrupole degrees
of freedom, i.e.  both $\beta$ and $\gamma$ deformations and the Euler
angles. This allows us to study in a consistent way an evolution of collective
properties  along the considered chain from spherical nuclei 
around the semi-magic \nuc{92}{Mo} to well deformed but soft
against $\gamma$ deformation heavier isotopes.
This approach was applied previously with a reasonable success for
several other regions of nuclei in.\cite{2004PR01,2006PR05,2007PR11,xx09pr01}

We should add that various properties of the discussed Mo
isotopes (of all or some of them) were studied in numerous papers using
various theoretical methods, let us mention only some of the most recent
ones.\cite{2009KV01,2009LA08,2006OZ01,2006ZA02,2006KO01}


\section{Theory and details of calculations}

The main tool of the collective part of the presented approach is the
general Bohr Hamiltonian, which is determined by 7 functions depending on
the deformation variables $\beta$, $\gamma$.  The functions are: the
potential energy and 6 inertial functions, including 3 moments of inertia 
(the inertial functions are frequently called mass parameters).
The microscopic part of our approach is
based on the Adiabatic Time Dependent HFB method which allows us to
calculate the seven mentioned functions starting from the effective
nucleon-nucleon interaction.  The deformation variables $\beta$, $\gamma$
are defined through mean values (in a HFB state) of the components of the
quadrupole tensor of mass distribution by $\beta\cos\gamma=D\langle
Q_{20}\rangle$, $\beta\cos\gamma=D\langle Q_{22}\rangle$ with
$D=\sqrt{\pi/5}/(A\overline{r^2})$.  Below we discuss the results obtained using
two well known variants of the 
Skyrme interaction: SIII\cite{1975be05} and Sly4.\cite{1997ch49} For the p-p
(pairing) channel we chose
the constant $G$ (also called state independent or of
the seniority type) interaction.  The strength of the pairing interaction
was determined
 by comparing  `experimental' 
pairing gaps with calculated  minimal quasi-particle energies for
\nuc{104-108}{Mo} isotopes. By `experimental' gap we mean a gap 
obtained from experimental mass differences with the use of the 5 point
formula.\cite{1988ma04} 
Technical details of the applied methods  can be found in.\cite{2004PR01}
Important aspect of our approach is that no free parameters are fitted to
obtain collective energies and B(E2) transition probabilities. Another important point
is that the so-called Thouless-Valatin corrections and  possible effects of the
pairing vibrations are simulated in a very rough way by multiplying the mass parameters by the
constant factor 1.3, for a deeper discussion of this point
see.\cite{xx09pr01}

\section{Results of calculations}

\subsection{Potential energy\label{sec:poten}}

Figs.~\ref{fig:pot086}-\ref{fig:pot108} show potential energy surfaces of
selected Mo isotopes calculated for both discussed variants of the
Skyrme interaction in the sextant
$\beta \leq 0.7$, $0^{\circ} \le \gamma \le 60^{\circ}$
of the deformation plane.  The plotted energy is relative to that of a
spherical shape of a given nucleus.  
Additional plot in each figure gives direct comparison of energies of axial shapes
for both versions of interaction.

It turns out that in the vicinity of the semi-magic \nuc{92}{Mo} both SIII
and SLy4 give almost identical potential energies, see e.g. 
Fig~\ref{fig:pot094}.  
For lighter, \nuc{84-86}{Mo}, and especially 
for heavier, \nuc{98-110}{Mo}, isotopes the energy from the SIII variant is
softer against deformation than the SLy4 energy.
One can be tempted to relate this different behaviour with increasing
neutron number to different global (i.e. concerning nuclear matter)
properties of these two interactions, especially the surface properties,
see e.g.\cite{2003BE12}. However, it is not clear
if such behaviour can be attributed to only one global 
property of a given interaction.
In the heavier region minima of the potential energy lie close (or on) the
prolate axis for the SIII and close to (or on) the oblate axis for the SLy4
version, see e.g.  Fig.~\ref{fig:pot108}.
It should be stressed, however, that in the approach applied in this paper
the most important is an overall shape of the potential energy and not a
precise location of its minimum.

\def\ryspot#1{
\includegraphics[scale=0.45]{s3_Mo#1.tr.v.sym.ps}
\includegraphics[scale=0.45]{sly4_Mo#1.h1.v.sym.ps}
\raisebox{3mm}{\includegraphics[scale=0.5]{vpot_ws_Mo#1.eps}}
}

\begin{figure}[htb]
\ryspot{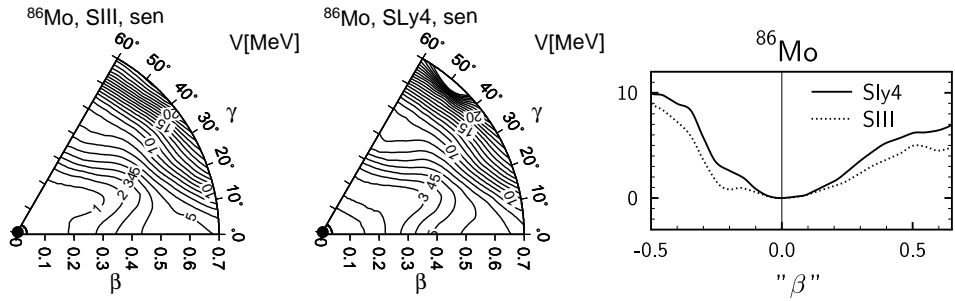}
\caption{\label{fig:pot086}Potential energy relative to that of a
spherical shape of the \nuc{86}{Mo} nucleus calculated using SIII and SLy4 versions
of the Skyrme interaction.}
\end{figure}

\begin{figure}[htb]
\ryspot{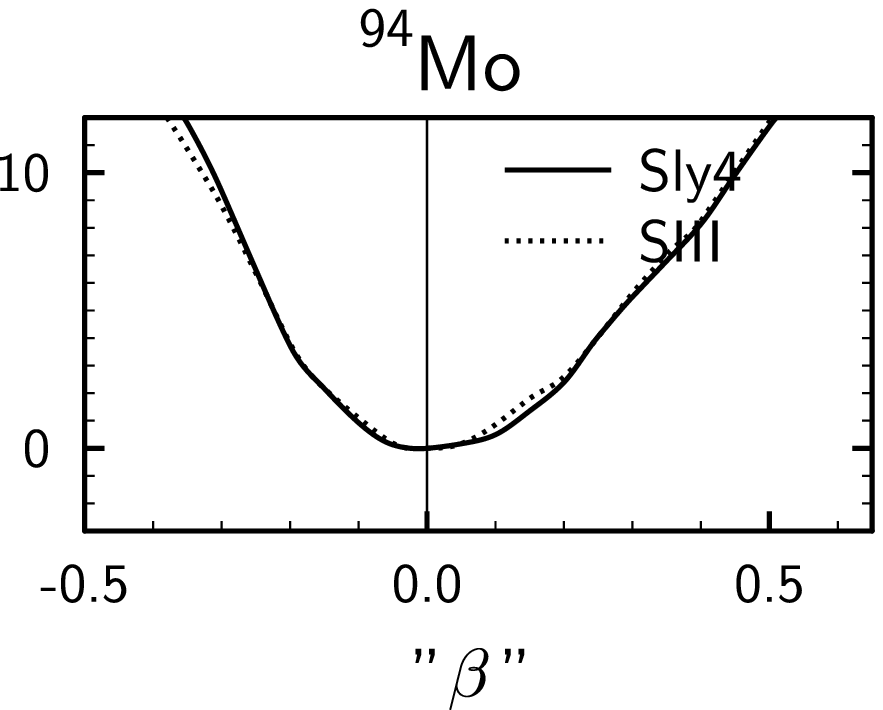}
\caption{\label{fig:pot094}Potential energy of the \nuc{94}{Mo} nucleus, see
also caption to Fig.~\ref{fig:pot086}.}
\end{figure}

\begin{figure}[htb]
\ryspot{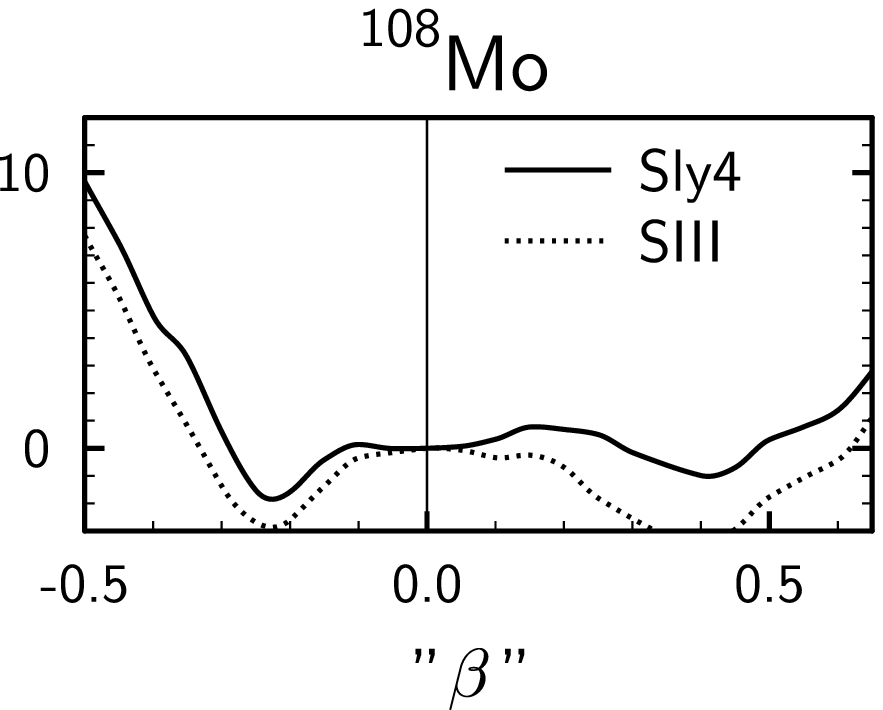}
\caption{\label{fig:pot108}Potential energy of the \nuc{108}{Mo} nucleus, see
also caption to Fig.~\ref{fig:pot086}.}
\end{figure}

The six inertial functions (mass parameters and moments of inertia) which
enter the Bohr Hamiltonian strongly depend on deformation but are not
plotted here for the sake of a limited space.
Preliminary analysis shows that differences in the collective spectra (see
Section 3.2) come
 from different potential energies rather than from different inertial
functions.

\subsection{Energy levels, E2 transitions}

Theoretical collective energy levels are obtained by solving 
an eigenproblem of the Bohr Hamiltonian. For comparison with experimental
data\cite{nndc0809} we chose  $2_1$ and $4_1$ states (from the ground state band) as well as
$2_2$ and $0_2$ states that can be considered as bandheads for the
quasi-$\gamma$ and quasi-$\beta$ bands. The results are presented in
Figs.~\ref{fig:lev21}-\ref{fig:lev02}.

\def\ryslev#1{
\includegraphics[scale=0.31]{s3_wlev#1_bw.eps}
\includegraphics[scale=0.31]{sly4_wlev#1_bw.eps}
}

\begin{figure}[htb]
\ryslev{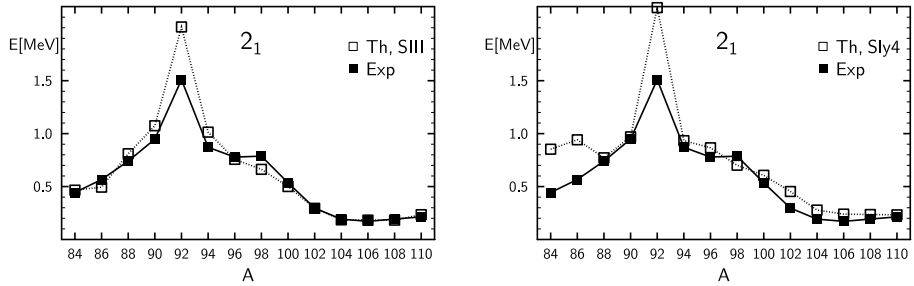}
\caption{\label{fig:lev21}Experimental and theoretical $2^+_1$ levels in the 
 the \nuc{84-110}{Mo} isotopes. Results for the SIII interaction are plotted in 
the left panel and for the Sly4 interaction --- in the right panel.}
\end{figure}

\begin{figure}[htb]
\ryslev{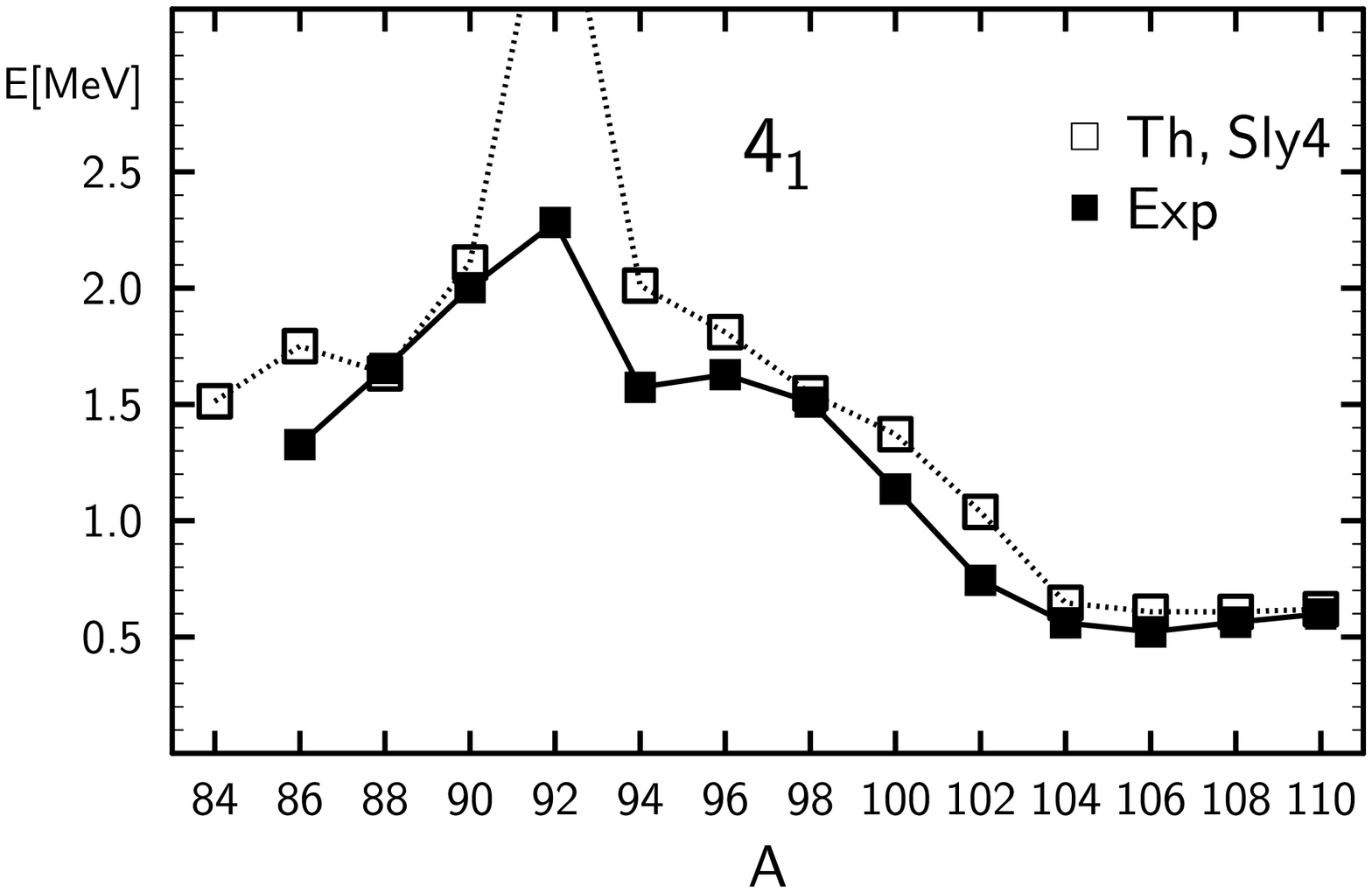}
\caption{\label{fig:lev41}Experimental and theoretical $4^+_1$ levels in the 
 the \nuc{84-110}{Mo} isotopes, see also caption to Fig.~\ref{fig:lev21}}
\end{figure}

\begin{figure}[htb]
\ryslev{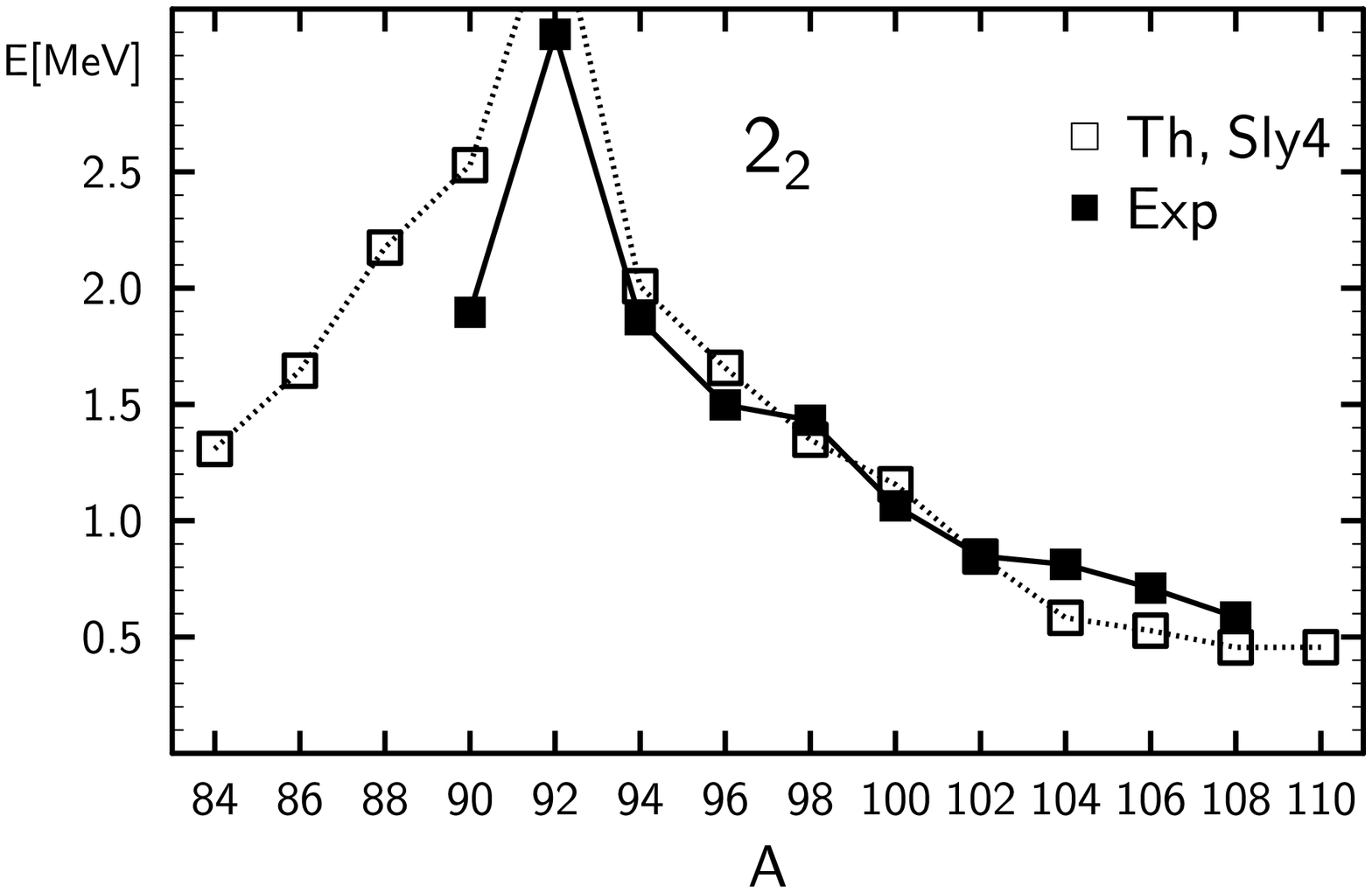}
\caption{\label{fig:lev22}Experimental and theoretical $2^+_2$ levels in the 
 the \nuc{84-110}{Mo} isotopes, see also caption to Fig.~\ref{fig:lev21}.}
\end{figure}

\begin{figure}[htb]
\ryslev{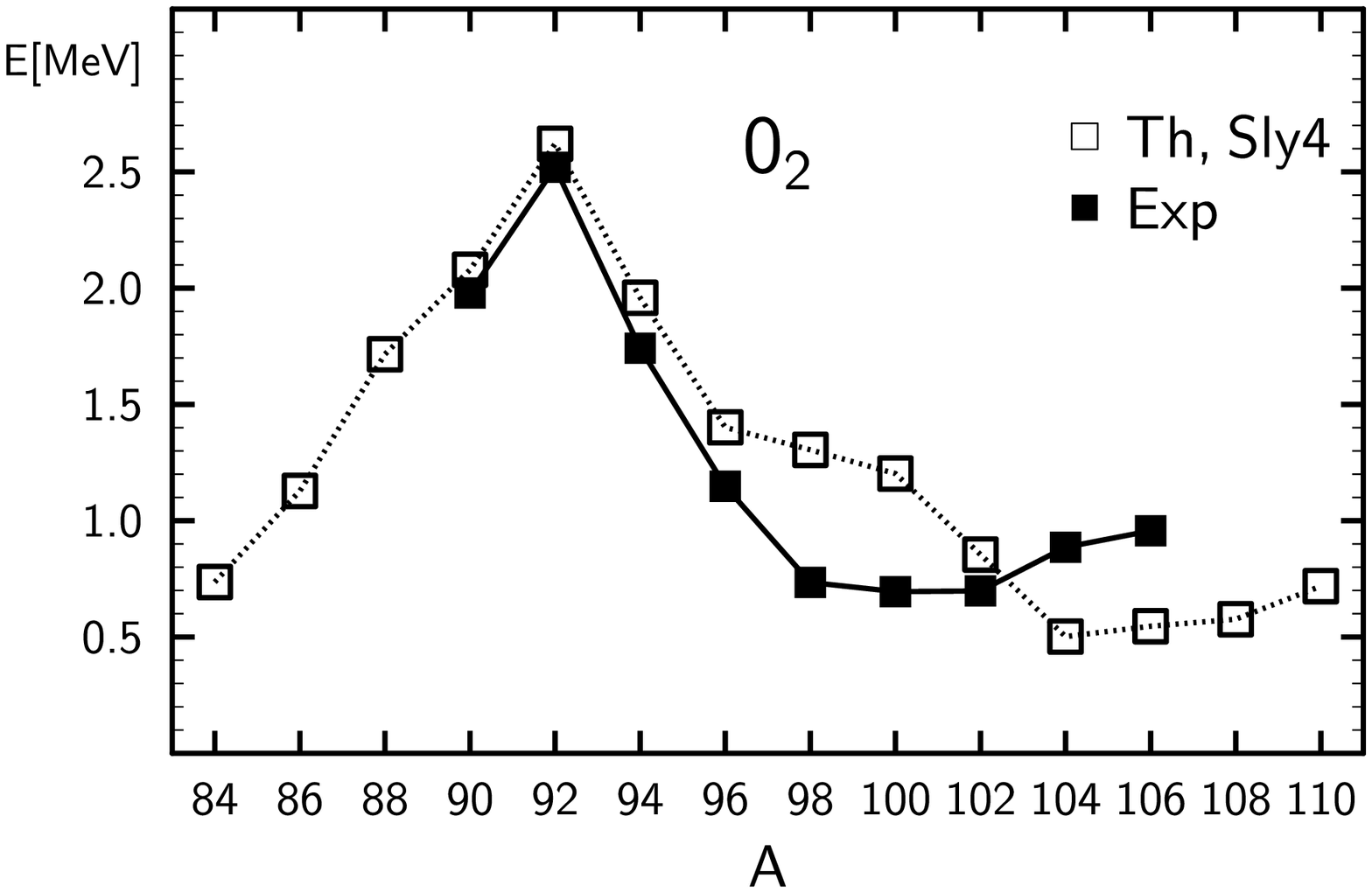}
\caption{\label{fig:lev02}Experimental and theoretical $0^+_2$ levels in the 
 the \nuc{84-110}{Mo} isotopes, see also caption to Fig.~\ref{fig:lev21}}
\end{figure}

Generally, both  SIII and SLy4 give similar, quite satisfactory, results,
however, it can be
clearly seen that differences in potential energies (see
Section~\ref{sec:poten})
lead to differences in collective energy levels.
The g.s. band states and $2_2$ levels are reproduced rather well,
with the exception of the case of the semi-magic \nuc{92}{Mo} nucleus. 
Problems with proper description of collective excitations in magic and, to
less extent, in semimagic nuclei are rather typical for approaches similar
to one employed in this paper (note that 
also in the case of the \nuc{92}{Mo} nucleus the mass parameters are multiplied by
the factor 1.3).
%
In the case of $0_2$ states there
are notable discrepancies 
between theory and experiment for the \nuc{96-100}{Mo} nuclei, however, more
detailed studies are needed because it is not clear if the first excited $0^+$
states, which lie very low 
in this nuclei (for \nuc{96}{Mo} even lower than $2^+_1$ level)
can  be  interpreted as collective quadrupole
excitations.\cite{2002ZI06,2006WR01}

Having determined theoretical collective wave functions one can calculate
E2 electromagnetic transition probabilities, which can give deeper  
insight into the nature of quadrupole collective states.
In our approach the collective E2 transition operator is given simply by a
quadrupole charge tensor, without any effective charges.
Fig.~\ref{fig:ee21} shows B(E2) reduced probabilities for the $2_1\rightarrow 0_1$ transitions 
within the ground state band.

General conclusion from the results of the present subsection is that 
the SIII version of the Skyrme interaction better reproduces experimental
data, for both energy levels and B(E2)'s, than the SLy4 version. However,
to judge the goodness of a particular variant of the interaction one should
consider also other observables not discussed here, such as absolute binding
energies, radii etc.

\def\rysee{
\includegraphics[scale=0.32]{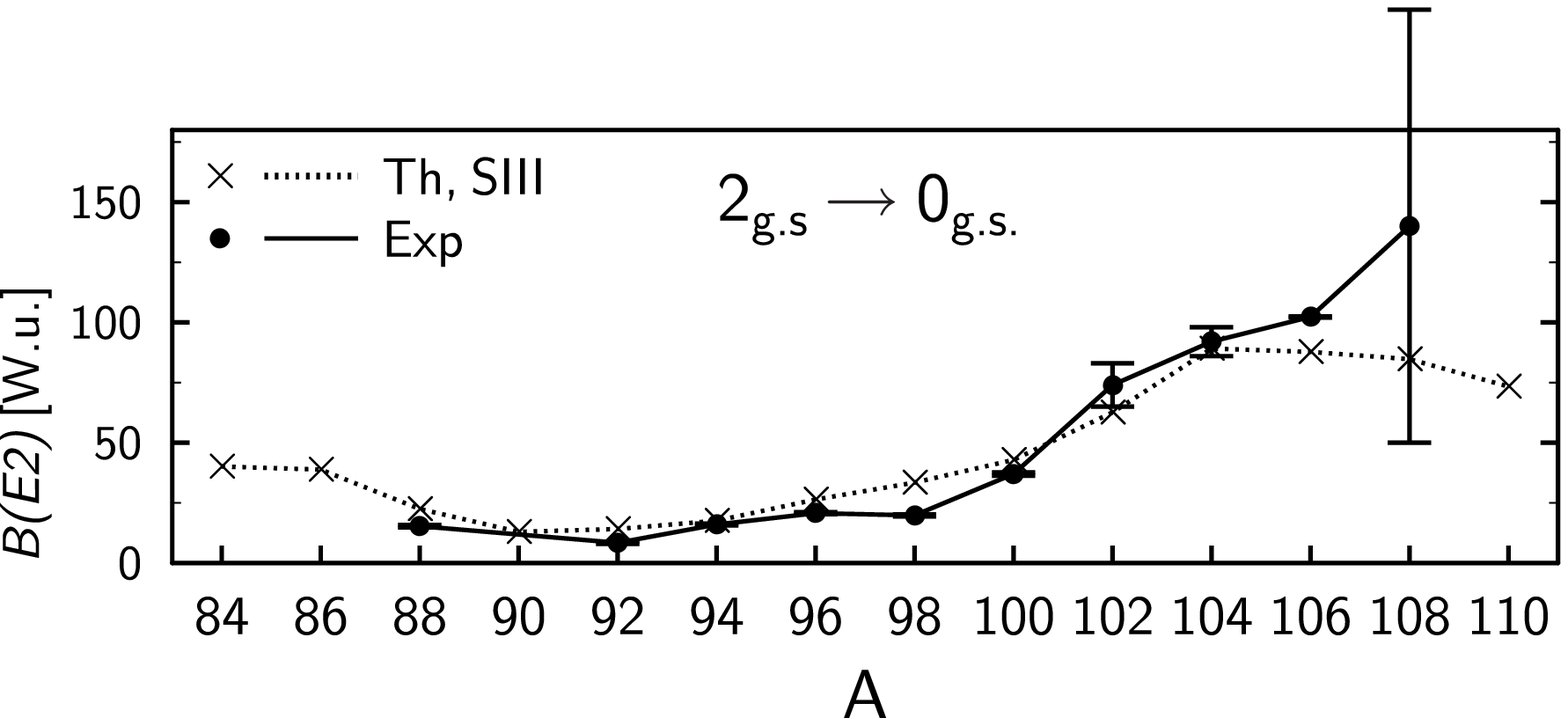}\hfil
\includegraphics[scale=0.32]{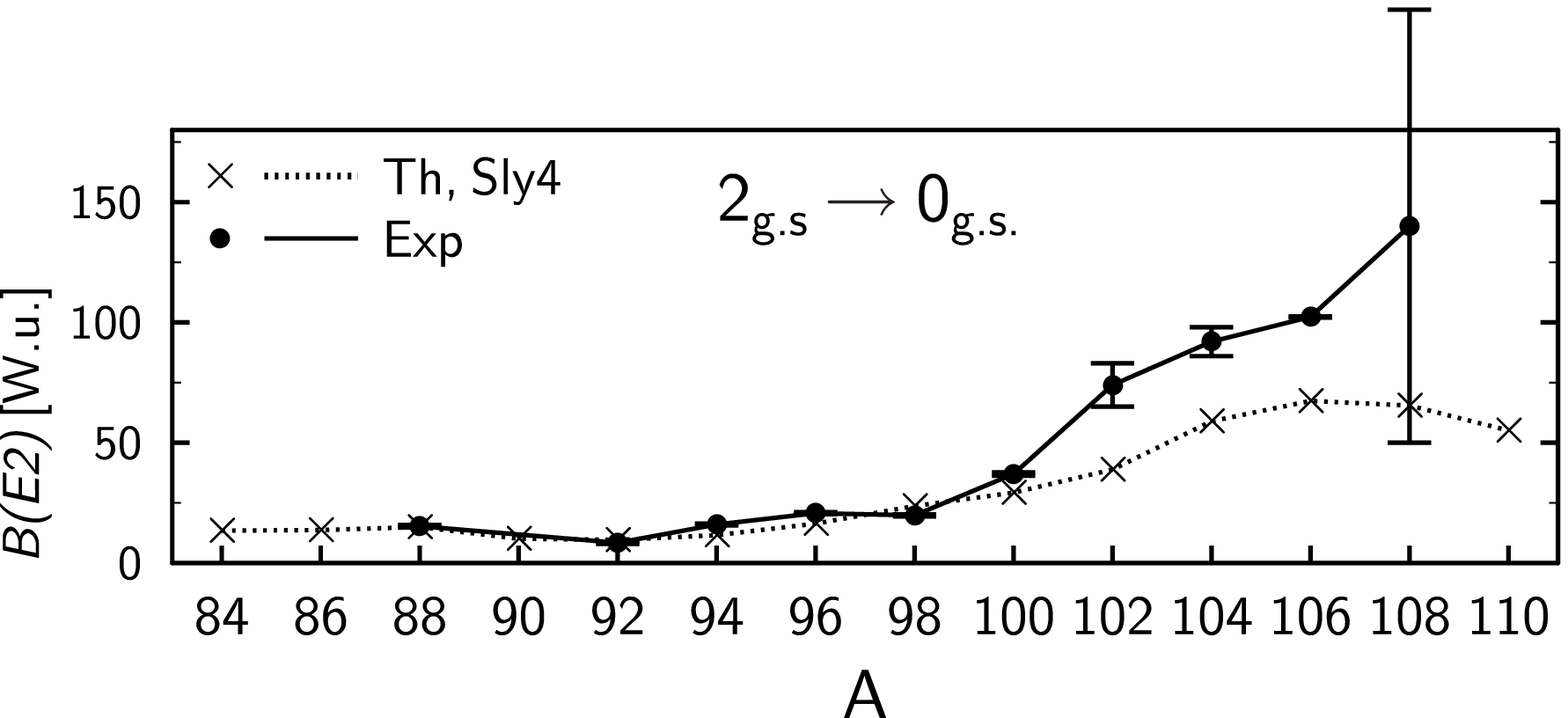}
}

\begin{figure}[htb]
\rysee
\caption{\label{fig:ee21}Comparison of experimental and theoretical B(E2)
values for the $2_1\rightarrow 0_{\rm g.s.}$ transition in the \nuc{84-110}{Mo}
isotopes. Left panel --- results for the SIII interaction, right panel SLy4.}
\end{figure}

\subsection{The \nuc{100}{Mo} nucleus}

\begin{figure}[htb]
\begin{center}
\fbox{\includegraphics[scale=0.52]{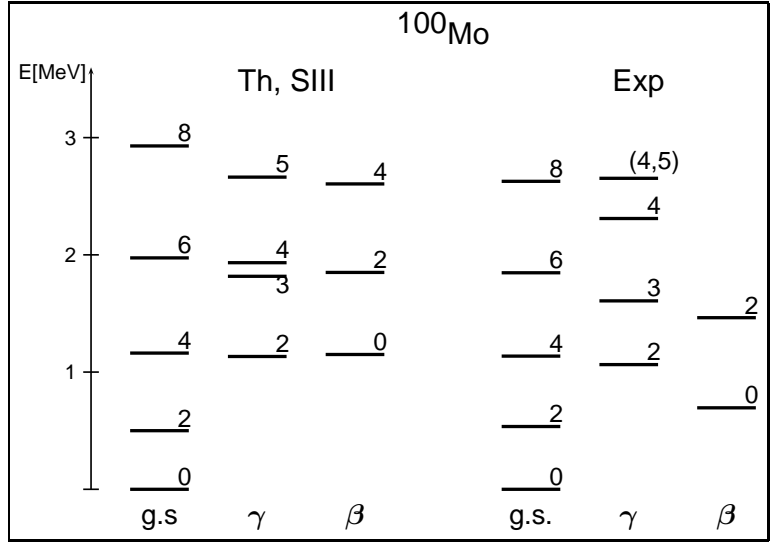}}
\end{center}
\caption{\label{fig:levmo100}Experimental and theoretical positive parity
levels of the \nuc{100}{Mo} nucleus.}
\end{figure}

Below we present more detailed results for the \nuc{100}{Mo} nucleus.
Fig.~\ref{fig:levmo100} shows the lowest energy levels of positive parity, grouped
tentatively into bands for the SIII variant (results for the Sly4 variant
do not differ significantly).
Qualitatively, the overall theoretical and
experimental patterns of energy levels are similar. However, position of the
$\beta$-band and details of the $\gamma$ band are not reproduced very well.

Table~\ref{mo100e2} contains experimental~\cite{nndc0809} and theoretical
values of the reduced probabilities for several E2 transitions in the
considered nucleus.

\def\subi#1{$_{#1}$}
\def\mcc#1#2{\multicolumn{#1}{c}{#2}}


\begin{table}[htb]
\tbl{\label{mo100e2}Experimental and theoretical B(E2) transition probabilities (in W.u.) for
the \nuc{100}{Mo} nucleus.}{%
\begin{tabular}{c@{\kern0mm}c@{\kern0mm}cr@{\kern0mm}rrr|c@{\kern0mm}c@{\kern0mm}cr@{\kern0mm}rrr}
\toprule
$J_i$ & & $J_f$ & \mcc{2}{Exp}  &\mcc{1}{SIII}& \mcc{1}{SLy4} &  $J_i$ & & $J_f$ & \mcc{2}{Exp}
&\mcc{1}{SIII}& \mcc{1}{SLy4}  \\ 
\colrule
 2\subi{1} &$ \rightarrow$ &  0\subi{1} &  37.0 & (0.7) & 42.9 &   29.2 &
 2\subi{2} &$ \rightarrow$ &  0\subi{2} &   5.5 & (0.8) & 12.1 &    3.0 \\ 
 4\subi{1} &$ \rightarrow$ &  2\subi{1} &    69 & (4) &  76.1 &  53.8 &
 2\subi{2} &$ \rightarrow$ &  2\subi{1} &    51 & (5) &  50.5 &  53.7 \\ 
 6\subi{1} &$ \rightarrow$ &  4\subi{1} &    94 & (14) &  102.7 &  75.6 &
 2\subi{2} &$ \rightarrow$ &  0\subi{1} &  0.62 & (0.05) &  0.8 &  0.06 \\ 
 8\subi{1} &$ \rightarrow$ &  6\subi{1} &   123 & (18) &  124.9 &  97.8 &
 2\subi{3} &$ \rightarrow$ &  2\subi{2} &    36 & (18) &  14.1 &   9.7 \\  
 0\subi{2} &$ \rightarrow$ &  2\subi{1} &    92 & (4) &  50.3 &  42.1 &
 2\subi{3} &$ \rightarrow$ &  0\subi{2} &    14 & (4) &  41.4 &  29.7 \\ 
 4\subi{2} &$ \rightarrow$ &  2\subi{2} &    30 & (6) &  45.4 & 39.5  &
 2\subi{3} &$ \rightarrow$ &  2\subi{1} &  0.28 & (0.08) &  0.003 &   0.1 \\ 
\botrule
\end{tabular}}
\end{table}

As one can see from Table~\ref{mo100e2}, the applied theoretical approach gives
good agreement with experiment for the \nuc{100}{Mo} nucleus, especially keeping in mind that there
are  no effective charges here. Again, the SIII results are slightly closer
to experiment than those of SLy4.

\section{Conclusions} 

The main conclusion is that the general Bohr Hamiltonian based on the ATDHFB
method with the Skyrme interaction describes general trends
in collective properties of the Mo isotopes reasonably well.  Moreover, as it can be seen at
least in the presented case of \nuc{100}{Mo}, this approach leads to quite
good qualitative agreement with  extensive experimental data for 
both energies and E2 transition probabilities.  To be more precise, we
showed
this for two (SIII and SLy4) among plenty of versions of the Skyrme
interactions.  In our opinion a study of collective properties can 
give an additional criterion when choosing the optimal variant of
interaction, or, more generally, the optimal shape of a density energy
functional. Such searches currently attract much attention.
However there are several questions which need more detailed studies.  Let
us mention some of them.  
Firstly, what are the effects of the quadrupole motion on the binding energy of
nuclei. To estimate such effects one should take into account the so called
zero point energy corrections given by the GCM method, see
e.g.\cite{2008KL04}
Secondly, the influence of the Thouless-Valatin terms and of 
the effects of the pairing vibrations on the collective kinetic energy is
simulated in a very rough way.\cite{1999LI38,2007PR07} More precise, 
quantitative estimation of such effects is still needed.

\end{document}